\documentclass{elsart}
\usepackage{graphicx,amssymb,amsmath}
\journal{Physics Letters A}
\begin{document}
\begin{frontmatter}
\title{Casimir force between dispersive magnetodielectrics}
\author{M. S. Toma\v s}
\address{Rudjer Bo\v skovi\' c Institute, P. O. B. 180, 10002
Zagreb, Croatia} \ead{tomas@thphys.irb.hr}

\begin{abstract} We extend our approach to the Casimir effect
between absorbing dielectric multilayers [M. S. Toma\v s, Phys.
Rev. A {\bf 66}, 052103 (2002)] to magnetodielectric systems. The
resulting expression for the force is used to numerically explore
the effect of the medium dispersion on the attractive/repulsive
force in a metal-magnetodielectric system described by the
Drude-Lorentz permittivities and permeabilities.
\end{abstract}

\end{frontmatter}

\section{Introduction}
The attractive theoretical possibility of obtaining the repulsive
Casimir force between (planar) magnetodielectrics \cite{Kenn02}
can hardly be realized at micron and submicron distances because
there is no natural medium with permeability different from unity
in the relevant frequency range \cite{Iann03}. Therefore, as
follows from the Lifshitz theory\cite{Lif0}, the Casimir force
should always be attractive at distances exploited in modern
experiments and applications \cite{Bord}. However, it has (also)
been conjectured that for media with nontrivial magnetic
properties the contributions from the low-frequency side of the
spectrum, where permeability dominates permittivity, may still
lead to the repulsive total force \cite{Kenn03}. In other words,
to obtain the Casimir force properly, it is necessary to account
for the medium dispersion over the entire spectrum and not only in
the $\lambda\sim d$ range \cite{Iann04}. In view of important
potential applications in the development of micro- and
nanoelectromechanical systems \cite{Buks} and nanotechnology in
general \cite{Bord}, it seems therefore worthwhile to consider the
effect of the medium dispersion on the Casimir force between
magnetodielectrics in more detail.

The aim of this work is to reconsider the statements made in Refs.
\cite{Kenn02,Iann03} and \cite{Kenn03} by analysing the
attractive/repulsive Casimir force at zero temperature in a very
simple metal-magnetodieletric system and accounting for the medium
dispersion. However, in order to rigorously provide an alternative
(and more familiar) result for the Casimir force to that given in
Ref. \cite{Kenn02}, we first extend the theory of the Casimir
effect between absorbing dielectric multilayers \cite{Tom02,Raa03}
to magnetodielectric systems.

\section{Theory}
Consider a multilayered system described by permittivity
$\varepsilon({\bf r},\omega)=\varepsilon'({\bf r},\omega)+
i\varepsilon''({\bf r},\omega)$ and permeability $\mu({\bf
r},\omega)=\mu'({\bf r},\omega)+i\mu''({\bf r},\omega)$ defined in
a  stepwise fashion, as depicted in Fig. 1.
\begin{figure}[htb]
\begin{center}
\resizebox{8cm}{!}{\includegraphics{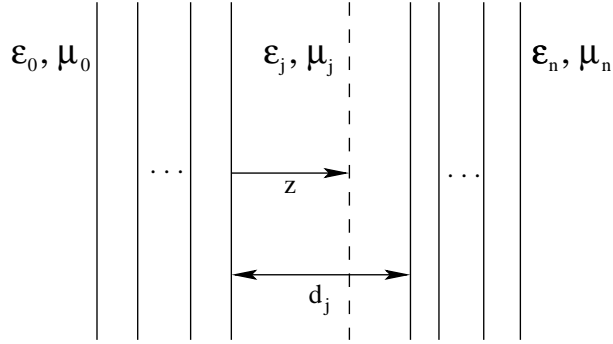}}
\end{center}
\caption{System considered schematically. The dashed line
represents the plane in a lossless layer where the stress  tensor
is calculated.}
\end{figure}

In order to calculate the relevant component of the Maxwell stress
tensor $T_{j,zz}$ \cite{LaLi}
\begin{equation}
T_{j,zz}=\frac{1}{8\pi}\left<E_zD_z-{\bf E}_\parallel\cdot{\bf
D}_\parallel+B_zH_z-{\bf B}_\parallel\cdot{\bf
H}_\parallel\right>_{{\bf r}\in(j)},
\label{T}
\end{equation}
we decompose the macroscopic field operators into positive
frequency and negative frequency parts according to
\begin{equation}
{\bf E}({\bf r},t)=\int_0^\infty d\omega {\bf E}({\bf
r},\omega)e^{-i\omega t}+\int_0^\infty d\omega {\bf E}^\dag({\bf
r},\omega)e^{i\omega t}.
\label{E}
\end{equation}
With the constitutive relations \cite{Dung03}
\begin{subequations}
\label{CE}
\begin{eqnarray}
{\bf D}({\bf r},\omega)&=& \varepsilon({\bf r},\omega){\bf E}({\bf
r},\omega)+
4\pi{\bf P}_N({\bf r},\omega),\\
{\bf H}({\bf r},\omega)&=&\frac{1}{\mu({\bf r},\omega)}{\bf
B}({\bf r},\omega)-4\pi{\bf M}_N({\bf r},\omega),
\end{eqnarray}
\end{subequations}
they obey the macroscopic Maxwell equations of the standard form.
Here ${\bf P}_N({\bf r},\omega)$ and ${\bf M}_N({\bf r},\omega)$
are the electric and magnetic noise polarization operators,
respectively, which obey the commutation rules (in the dyadic
form):
\begin{subequations} \label{CR}
\begin{equation}
[{\bf P}_N({\bf r},\omega),{\bf P}^\dagger_N({\bf r}',\omega')]
=\frac{\hbar}{4\pi^2}\varepsilon''({\bf r},\omega) {\bf
I}\delta({\bf r}-{\bf r}')\delta(\omega-\omega'),
\end{equation}
\begin{equation}
[{\bf M}_N({\bf r},\omega),{\bf M}^\dagger_N({\bf r}',\omega')]
=\frac{\hbar}{4\pi^2} \frac{\mu''({\bf r},\omega)}{|\mu({\bf
r},\omega)|^2} {\bf I}\delta({\bf r}-{\bf r}')
\delta(\omega-\omega'),
\end{equation}
\end{subequations}
where ${\bf I}$ is the unit dyadic. Therefore, any (annihilation)
field operator is related to ${\bf P}_N({\bf r},\omega)$ and ${\bf
M}_N({\bf r},\omega)$ via the classical Green function ${\bf
G}({\bf r},{\bf r'};\omega)$ satisfying
\begin{equation}
\left[\nabla\times\frac{1}{\mu({\bf r},\omega)}\nabla\times-
\varepsilon({\bf r},\omega)\frac{\omega^2}{c^2}{\bf
I}\cdot\right]{\bf G}({\bf r},{\bf r'};\omega)=4\pi{\bf I}
\delta({\bf r}-{\bf r'})
\label{GF}
\end{equation}
and the outgoing-wave condition at infinity. As a consequence, all
field correlation functions can be expressed through the Green
function in accordance with the fluctuation-dissipation theorem
\cite{LiPi}. In particular, for the electric-field correlation
function we find
\begin{equation}
\left<{\bf E}({\bf r},\omega){\bf E}^\dagger({\bf
r}',\omega')\right> =\frac{\hbar}{\pi}\frac{\omega^2}{c^2} {\rm
Im}{\bf G}({\bf r},{\bf r'};\omega)\delta(\omega-\omega'),
\label{EE}
\end{equation}
and the magnetic-field correlation function is obtained from this
expression using ${\bf B}({\bf r},\omega)=
(-ic/\omega)\nabla\times{\bf E}({\bf r},\omega)$.

Applying the above results to the $j$th layer and taking into
account that $\varepsilon_j(\omega)$ and $\mu_j(\omega)$ are real
and that ${\bf P}_N({\bf r},\omega)=0$ and ${\bf M}_N({\bf
r},\omega)=0$ in this region, for the relevant correlation
functions in Eq. (\ref{T}) we find
\begin{subequations}
\label{CF}
\begin{equation}
\left<{\bf E}({\bf r},t){\bf D}({\bf r},t)\right>_{{\bf r}\in(j)}
 =\frac{\hbar}{\pi}\int_0^\infty d\omega \varepsilon_j(\omega)
 \frac{\omega^2}{c^2}{\rm Im}{\bf G}_j({\bf r},{\bf
r};\omega), \label{ED}
\end{equation}
\begin{equation}
\left<{\bf B}({\bf r},t){\bf H}({\bf r},t)\right>_{{\bf r}\in(j)}
=\frac{\hbar}{\pi}\int_0^\infty d\omega\frac{1}{\mu_j(\omega)}
{\rm Im}{\bf G}^B_j({\bf r},{\bf r};\omega). \label{BH}
\end{equation}
\end{subequations}
Here ${\bf G}_j({\bf r},{\bf r'};\omega)$ is the Green  function
element for ${\bf r}$ and ${\bf r'}$ both in the layer $j$, and
\begin{equation}
{\bf G}^B_j({\bf r},{\bf r'};\omega)= \nabla\times{\bf G}_j({\bf
r},{\bf r'};\omega)\times \stackrel{\leftarrow}{\nabla'}
\label{GB}
\end{equation}
is the corresponding Green function element for the magnetic
field.

The Casimir force per unit area $f_j$  between the stacks bounding
the $j$th layer equals to $T_{j,zz}$ when omitting its
infinite-medium part \cite{Raa05}. Formally, this is done by
replacing the Green function in Eq. (\ref{CF}) with its scattering
part
\begin{equation}
{\bf G}^{\rm sc}_j({\bf r},{\bf r'};\omega)= {\bf G}_j({\bf
r},{\bf r'};\omega)-{\bf G}^0_j({\bf r},{\bf r'};\omega),
\label{Gsc}
\end{equation}
where ${\bf G}^0_j({\bf r},{\bf r'};\omega)$ is the
infinite-medium Green function. Writing ${\bf G}^{\rm sc}_j({\bf
r},{\bf r'};\omega)$ simultaneously as
\begin{equation}
{\bf G}^{\rm sc}_j({\bf r},{\bf r'};\omega)=
\mu_j(\omega){\underline{\bf G}}^{\rm sc}_j({\bf r},{\bf
r'};\omega),
\end{equation}
we have
\begin{eqnarray}
\label{fj} f_j&=&\frac{\hbar}{4\pi}{\rm Im}\int_0^\infty
\frac{d\omega}{2\pi}\left\{k^2_j(\omega) \left[\underline{G}^{\rm
sc}_{j,zz}({\bf r},{\bf r};\omega)- \underline{G}^{\rm
sc}_{j,\parallel}({\bf r},{\bf r};\omega)\right]
\right.\nonumber\\
&&\left.+\underline{G}^{B,\;{\rm sc}}_{j,zz}({\bf r},{\bf
r};\omega)- \underline{G}^{B,\;{\rm sc}}_{j,\parallel}({\bf
r},{\bf r};\omega\right\},
\end{eqnarray}
where
\begin{equation}
k_j(\omega)=\sqrt{\varepsilon_j(\omega)\mu_j(\omega)}\frac{\omega}{c}
\label{kj}
\end{equation}
is the wave vector in the layer and $G^{\rm sc}_{j,\parallel}({\bf
r},{\bf r}';\omega)=G^{\rm sc}_{j,xx}({\bf r},{\bf
r}';\omega)+G^{\rm sc}_{j,yy}({\bf r},{\bf r}';\omega)$.

In terms of ${\underline{\bf G}}^{sc}_j({\bf r},{\bf r'};\omega)$,
the above expression for the force formally coincides with the
corresponding result for a purely dielectric multilayer
\cite{Tom02}. As follows from Eq. (\ref{GF}), ${\underline{\bf
G}}^{sc}_j({\bf r},{\bf r'};\omega)$ itself is of the same form as
the Green function ${\bf G}^{sc}_j({\bf r},{\bf r'};\omega)$ for a
purely dielectric system \cite{Tom95}, the only difference being
that the wave vectors in the layers are now given according to Eq.
(\ref{kj}) and the Fresnel coefficients of the surrounding stacks
are modified because of the different magnetic properties of the
system in the present case. Accordingly, with these modifications
in mind, we can formally adopt all subsequent results obtained in
Ref. \cite{Tom02}. Thus, with
$\varepsilon_j(\omega)=\mu_j(\omega)=1$, according to Eq. (2.15)
of Ref. \cite{Tom02} we have for the Casimir force
\begin{eqnarray}
f_j&=&-\frac{\hbar}{\pi}{\rm Re}\int_0^\infty d\omega
\int\frac{d^2{\bf k}}{(2\pi)^2}\beta_j
\sum_{q=p,s}\frac{1-D_{qj}(\omega,k)}{D_{qj}(\omega,k)}\nonumber\\
&=&\frac{\hbar}{2\pi^2}\int_0^\infty d\xi \int^\infty_0
dkk\kappa_j \sum_{q=p,s}\frac{1-D_{qj}(i\xi,k)}{D_{qj}(i\xi,k)},
\label{fjf}
\end{eqnarray}
where $\beta_j(\omega,k)=\sqrt{\omega^2/c^2-k^2}$,
\begin{equation}
D_{qj}(\omega,k)=1-r^q_{j-}(\omega,k) r^q_{j+}(\omega,k)
e^{2i\beta_j d_j},
\label{Dj}
\end{equation}
and $r^q_{j\pm}(\omega,k)$ are the reflection coefficients of the
right ($+$) and left ($-$) stack for TE ($q=s$)- and TM
($q=p$)-polarized waves, respectively. The second line in Eq.
(\ref{fjf}) is obtained in the usual way by converting the
integral over the real $\omega$-axis to that along the imaginary
$\omega$-axis, letting $\omega=i\xi$, introducing
$\kappa_j(\xi,k)\equiv -i\beta_j(i\xi,k)=\sqrt{\xi^2/c^2+k^2}$,
and noticing the reality of the integrand.

\section{Discussion: Repulsive force}
For numerical calculations, it is convenient to make
transition to polar coordinates in the second line of Eq.
(\ref{fjf}). Letting $\xi/c=\kappa_j\cos\phi$,
$k=\kappa_j\sin\phi$, and then $x=2\kappa_j d$, we have (hereafter
 we omit the index $j$)
\begin{eqnarray} f&=&f_{\rm id}(d)\frac{15}{2\pi^4}\int_0^\infty dxx^3e^{-x}
\int^{\pi/2}_0 d\phi\sin\phi \nonumber\\
&&\times\sum_{q=p,s}\frac{r^q_-(x,\phi) r^q_+(x,\phi)}
{1-r^q_-(x,\phi) r^q_+(x,\phi)e^{-x}}, \label{f}
\end{eqnarray}
where $f_{\rm id}(d)=\pi^2\hbar c/240d^4$ is the Casimir force in
the ideal (attraction) case \cite{Cas} and $r^q_\pm(x,\phi)=
r^q_\pm(i\frac{cx}{2d}\cos\phi,\frac{x}{2d}\sin\phi)$. For a
vacuum-medium interface we explicitly have
\begin{equation}
\label{rps} r^p(x,\phi)=
\frac{\varepsilon-\sqrt{\varepsilon\mu\cos^2\phi+\sin^2\phi}}
{\varepsilon+\sqrt{\varepsilon\mu\cos^2\phi+\sin^2\phi}},\;\;\;
r^s(x,\phi)=r^p[\varepsilon\leftrightarrow\mu],
\end{equation}
where $\varepsilon$ and $\mu$ are functions of
$\omega=i(cx/2d)\cos\phi$. Clearly, owing to this dependence of
$\varepsilon$ and $\mu$ on $d$, the integral in Eq. (\ref{f}) is
also a (slowly varying) function of the separation between the
slabs.

Equations (\ref{f}) and (\ref{rps}) are equivalent to Eqs. (4) and
(5) quoted by Kenneth {\it et al.}\cite{Kenn02}. In the subsequent
lines, these authors proceeded by regarding $\varepsilon$ and
$\mu$ as some constants (appropriate for the relevant frequency
range \cite{Kenn03}). In the following, we reconsider their
analysis adopting the single-oscillator model for the electric
($e$) and magnetic ($m$) polarization \cite{Rupp}. Each slab is
therefore described by permittivity and permeability of the
Drude-Lorentz type
\begin{equation}
\label{emu} \{\varepsilon,\mu\}=1+\frac{(2\omega_{P\nu}d/c)^2}
{(2\omega_{T\nu}d/c)^2 +x\cos\phi(x\cos\phi+ 2\gamma_\nu d/c)},
\end{equation}
where $\omega_{P\nu}$ ($\nu=e,m$) measure the coupling strengths
between the medium and the electromagnetic field, whereas
$\omega_{T\nu}$ and $\gamma_\nu$ are resonant frequencies of the
medium and linewidths of the associated resonances, respectively.
It is immediately seen that the non-dispersive regime is reached
at distances $\omega_{T\nu}d/c\gg 1$, where $\varepsilon$ and
$\mu$ acquire their static values
$\{\varepsilon,\mu\}(0)=1+\omega^2_{P\nu}/\omega^2_{T\nu}$. For
convenient measures of the magnitudes of $\varepsilon$ and $\mu$,
we can therefore take the quantities
$\omega_{P\nu}/\omega_{T\nu}$. To have a feeling about the
importance of the second term in the denominator of Eq.
(\ref{emu}), we note that the main contribution to the integral in
Eq. (\ref{f}) comes from the value of $x$ of about $2$.

We have performed numerical calculations for a
metal-magnetodielectric system. For the dielectric function
$\varepsilon_M$ of the metal we set $\omega_{Pe}=\omega_P$,
$\omega_{Te}=0$, and $\gamma_e=\gamma$ in Eq. (\ref{emu}) and for
the permeability we set $\mu_M=1$ ($\omega_{Pm}=0$). To simulate
$\gamma$ of good metals (such as Au, Cu, Al) at zero temperature,
we have let $\gamma=10^{-6}\gamma_{\rm Au}(300\;{\rm K})$ which,
in conjunction with $\hbar\gamma_{\rm Au}(300\;{\rm K})=35$ meV
\cite{Lam}, corresponds to the residual relaxation parameter of
gold at $T=0$ K \cite{Bez}. The metal is considered a reference
medium and characteristic frequencies of the magnetodielectric are
scaled with $\omega_P$. Also, the distance between the slabs is
measured in units $k^{-1}_P=c/\omega_P$. For orientation, we note
that in the case of Au ($\hbar\omega_P=9.0$ eV \cite{Lam}),
$k^{-1}_P\simeq 22$ nm and a similar value is found for other good
metals. Thus, submicron distances correspond to $k_P d\lesssim
50$.
\begin{figure}[htb]
\begin{center}
\resizebox{8cm}{!}{\includegraphics{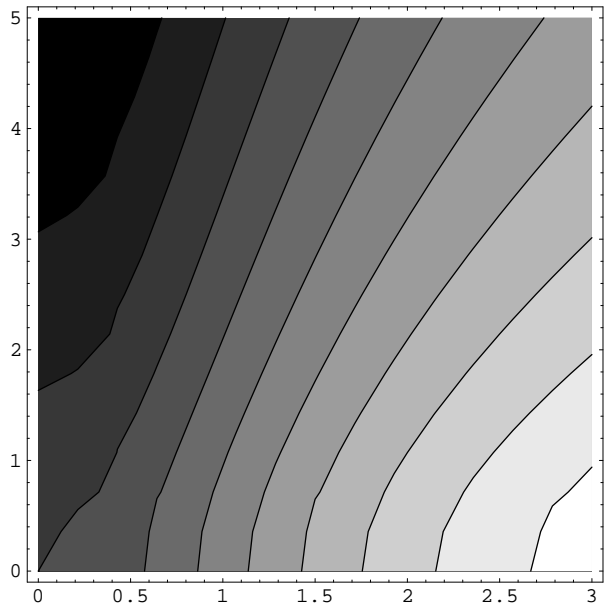}}
\end{center}
\caption{Contour plot of $f/f_{\rm id}$ as a function of
$P_e=\omega_{Pe}/\omega_{Te}$ and $P_m=\omega_{Pm}/\omega_{Tm}$
for $k_P d=1$, $\omega_{Te}/\omega_P=0.7$, and
$\omega_{Tm}/\omega_P=0.5$. Damping parameters are
$\gamma=3.9\times 10^{-9}\omega_P$ and
$\gamma_e/\omega_{Te}=\gamma_m/\omega_{Tm}=10^{-2}$ (these
parameters are kept fixed in all simulations). Constant-force
lines are plotted in steps of $\Delta f/f_{\rm id}=0.02$, starting
from $f/f_{\rm id}=-0.06$ at the top left corner. \label{PP}}
\end{figure}

The dependence of the reduction factor $f/f_{\rm id}$ on plasma
frequencies $\omega_{P\nu}$ (since $\omega_{T\nu}$ are fixed) of
the magnetodielectric is illustrated in Fig. \ref{PP}. Note that
for this $d$, the figure actually represents the Casimir force in
units $f_0=\pi^2\hbar ck_P^4/240$ ($\simeq 5.67\; {\rm kN}/{\rm
m}^2$ for Au). One observes that the transition from the
attractive to repulsive regime occurs roughly when $P_m>P_e$ and
that the constant-force lines at larger $P_m$ and $P_e$ in the
transition region  are (nearly) straight lines characterized by a
(nearly) constant ratio $P_m/P_e$. Similar behavior, however, in
terms of $\varepsilon$ and $\mu$, is found for the border line
between the attractive and repulsive regions in Fig. 1a of
Ref.\cite{Kenn02}. We also note that our simulation agrees with
the conclusion that for large $\varepsilon$ and $\mu$ the sign and
the magnitude of the force depend on the surface impedance
$Z=\sqrt{\mu/\varepsilon}$ of the magnetodielectric \cite{Kenn02}.
This can be seen if we rewrite Eq. (\ref{emu}) in terms of the
relative quantities $P_\nu$, $Q_\nu=\omega_{T\nu}/\omega_P$ and
$\tilde{d}=k_Pd$
\begin{equation}\label{emPQ}
\{\varepsilon,\mu\}=1+\frac{P^2_\nu  (2Q_\nu\tilde{d})^2}
{(2Q_\nu\tilde{d})^2 +x\cos\phi(x\cos\phi+ 2Q_\nu\tilde{d}
\gamma_\nu/\omega_{T\nu})}.
\end{equation}
At large $P_\nu$ one can disregard unity here, so that $Z\sim
P_m/P_e$. Accordingly, (nearly) straight lines seen in the
($P_m,P_e$) contour plot correspond to constant-$Z$ lines. Since
for this separation between the slabs $\varepsilon$ and $\mu$ are
strongly dispersive (note that $k_{Te}d=0.7$ and $k_{Tm}d=0.5$),
one must conclude that accounting for the medium dispersion does
not much affect the conclusions drawn in Ref. \cite{Kenn02}
regarding the ($\varepsilon,\mu$)-dependence of the force.

\begin{figure}[htb]
\begin{center}
\resizebox{8cm}{!}{\includegraphics{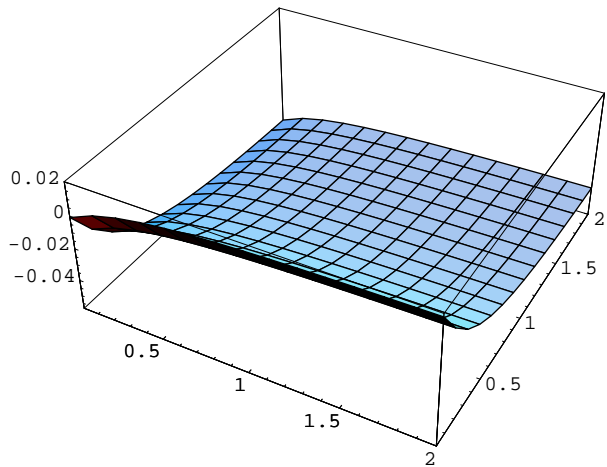}}
\end{center}
\caption{$f/f_{\rm id}$ as a function of $\omega_{Te}/\omega_P$
and $\omega_{Tm}/\omega_P$ for $k_P d=1$,
$\omega_{Pe}/\omega_{Te}=0.5$, and
$\omega_{Pm}/\omega_{Tm}=3$.\label{QQ}}
\end{figure}

Clearly, of crucial importance for the appearance of the repulsive
force  is the relative position of the electric and magnetic
resonance of the magnetodielectric. Figure \ref{QQ} illustrates
the variation of $f/f_{\rm id}$ with transverse frequencies
($Q_\nu$) of the magnetodielectric while keeping the magnitudes of
$\varepsilon$ and $\mu$ (determined by $P_\nu$) fixed. At large
$Q_\nu$ the force tends to a constant value because
$\{\varepsilon,\mu\}\rightarrow 1+P^2_\nu$ in Eq. (\ref{emPQ}). Of
course, in the opposite limit, the force vanishes as
$\{\varepsilon,\mu\}\simeq 1$ for small $Q_\nu$. It is seen that
the appearance and magnitude of the repulsive force is governed by
the position of the magnetic resonance for almost all $Q_e$. Note
that in the region where $Q_e>Q_m$, which is normally always the
case, the force is attractive at this distance.

\begin{figure}[h]
\begin{center}
\resizebox{8cm}{!}{\includegraphics{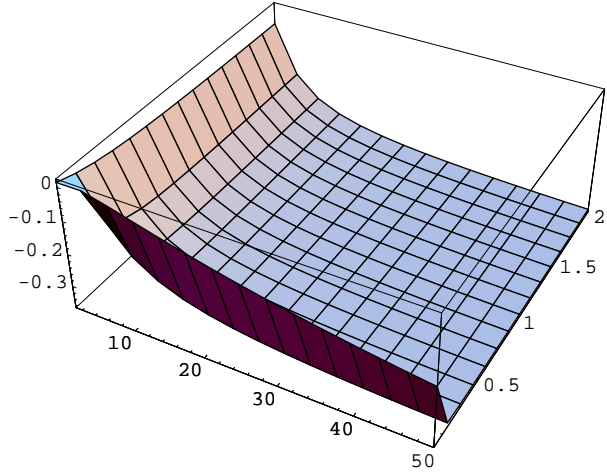}}
\end{center}
\caption{$f/f_{\rm id}$ as a function of the distance $k_P d$
between the slabs and the position $\omega_{Tm}/\omega_P$ of the
magnetic resonance for $\omega_{Pe}/\omega_{Te}=0.5$,
$\omega_{Te}/\omega_P=0.7$, and $\omega_{Pm}/\omega_{Tm}=3$.
\label{dQ}}
\end{figure}

\begin{figure}[htb]
\begin{center}
\resizebox{8cm}{!}{\includegraphics{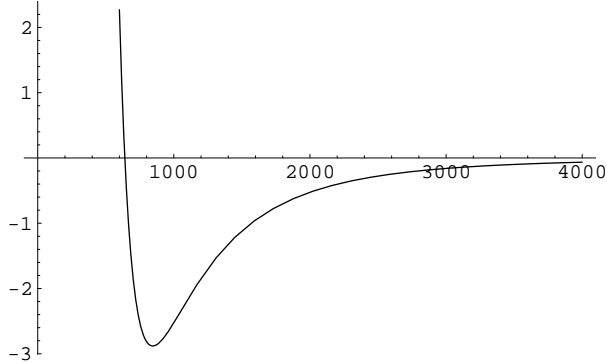}}
\end{center}
\caption{$f$ in units $10^{-14}f_0$ as a function of the distance
$k_P d$ between the slabs for a "realistic" system:
$\omega_{Te}/\omega_P=0.1$, $\omega_{Tm}/\omega_P=10^{-4}$ and
other parameters are the same as in Fig. \ref{dQ}. \label{real}}
\end{figure}

Since $Q_\nu$ and $\tilde{d}$ appear in the form of the product
$Q_\nu\tilde{d}=\omega_{T\nu}d/c$ in Eq. (\ref{emPQ}), the
mismatch between the electric and magnetic resonances can be
compensated by properly adjusting $d$ and thus one can in
principle still observe the repulsive force although at larger
distances. This possibility is illustrated in Fig. \ref{dQ}. One
observes that, indeed, even for very small values of $Q_m$ ($\ll
Q_e$) at larger distances there is a repulsive force between the
media. To estimate the distance at which the repulsive force could
be observed in a (possibly) realistic system,  we note that the
highest frequencies known so far at which permeabilities differ
significantly from unity are (for some antiferromagnetics) in the
THz range  \cite{Kli}. Realistic values of the parameters $Q_\nu$
are therefore $Q_e\sim 0.1$, corresponding to $\omega_{Te}$ in the
range of optical frequencies, and $Q_m\sim 10^{-4}$, corresponding
to $\omega_{Tm}$ in the THz range. The distance dependence of the
(total) force for these parameters across the repulsive region is
presented in Fig.\ref{real}.

As seen, the force is attractive at small distances, which
supports claims made in Ref \cite{Iann03}. With increasing $d$,
however, the effect of permeability is enforced and the force
crosses over to the repulsive regime already at $k_Pd\simeq 640$
(which is $\simeq 14 {\rm \mu m}$ for Au). Thus, the effect of
magnetic polarization appears at distances smaller by two orders
of magnitude than it would be expected on the basis of a simple
estimate $d\sim\lambda_{Tm}=2\pi\times 10^4\;k^{-1}_P$. Evidently,
this effect is due to the cumulative contribution of the waves
from the low-frequency side of the spectrum and is therefore a
direct consequence of the medium dispersion at long wavelengths.
To estimate the magnitude of the repulsive force at such
distances, we calculate its largest value occurring at $k_Pd\simeq
850$, where we find $f\simeq -2.88\times 10^{-14}f_0$.
Accordingly, the maximal repulsive force between, e.g. Au and this
hypothetical [$\omega_{Te}=0.9$ eV, $\omega_{Pe}=0.45$ eV,
$\gamma_e=9$ meV, $\omega_{Tm}=0.9$ meV, $\omega_{Pm}=2.7$ meV,
and $\gamma_m=9 \;\mu{\rm eV}$] magnetodielectric is $f\simeq
-0.16$ ${\rm nN}/{\rm m}^2$ and is observed at the distance
$d\simeq 18.7 \;{\rm \mu m}$.

Of course, at higher temperatures the above estimate becomes
invalid as at such large separations between the slabs the thermal
corrections must be accounted for. The force at nonzero
temperature is formally obtained from Eq. (\ref{fjf}) by letting
\cite{Raa03} $\xi\rightarrow\xi_m=2\pi m k_BT/\hbar$ and
$\int_0^\infty d\xi\rightarrow (2\pi
k_BT/\hbar)\sum_{m=0}^\infty(1-\delta_{m0}/2)$, where $\xi_m$ are
the Matsubara frequencies. One may immediately conclude that the
force in the present system is always attractive at room
temperature. Indeed, at $T=300$ K and for Au, the $m$th
contribution to the force is governed by the quantity
\[\frac{\xi_md}{c}=2\pi m\frac{k_BT}{\hbar\omega_P}k_Pd=
1.8\times 10^{-2} m\;k_Pd.\] Thus, at distances $k_Pd\gtrsim 10^2$
only the $m=0$ term contributes as all other terms are strongly
exponentially damped. Since for the Drude metals $r^s_M(0,k)=0$
\cite{com}, this term depends only on the dielectric properties of
the magnetodielectric and the force is therefore attractive. We
have verified this by directly calculating the force at $T=300$ K
and, for example, at $k_Pd\simeq 850$, where the zero-temperature
force is maximally repulsive, we have obtained $f\simeq 5.09\times
10^{-13}f_0\simeq 2.89$ ${\rm nN}/{\rm m}^2$.

The validity of the above estimates rests upon the adequacy of the
Drude (local) model for the metal adopted. This model is often
used in the considerations and precise calculations of the Casimir
force between metals and is generally believed to adequately
describe the properties of good metals at low frequencies
\cite{Lam,Brev}. Strictly speaking, however, the electromagnetic
response of a metal is nonlocal, at least in the frequency range
of the anomalous skin effect, and should therefore be described by
the nonlocal reflection coefficients $r^q_M$ \cite{LaLi}. The
effect of the metal nonlocality below $\omega<\omega_P$ on the
Casimir force between two metal (Au) plates at zero temperature
has recently been considered in detail in Ref. \cite{Esq} and it
has been found that anomalous skin effect produces only a minor
correction (below 0.5\% for $d$ above $50$ nm) to the force
calculated using the local description of the metal (which also
agrees with the assertion given in Ref. \cite {Brev}). Moreover,
as can be concluded from Fig. 9 of Ref. \cite{Esq}, this
correction diminishes and saturates with increasing distance
between the plates. Accordingly, accounting for the anomalous skin
effect of the metal cannot significantly change the above
estimates on the repulsive force at $T=0$ K. As follows from the
small frequency behaviour of the surface impedances \cite{Esq},
since the property $r^s_M(0,k)=0$ holds nonlocally as well, the
force at $T=300$ K is, as before, always attractive. However,
since also $r^p_M(0,k)\neq 1$, contrary to the generally accepted
$r^p_M(0,k)=1$, further consideration of the effect of the metal
nonlocality on the Casimir force at finite temperatures is needed.

We end this short discussion by noting that very recently Henkel
and Joulain reported a similar analysis \cite{HeJou}, however,
paying more attention to the dependence of the force on
geometrical rather than on material parameters as we have done. In
this respect, these two works are complementary. As concerns our
findings on the distance dependence of the force, these are
(generally) in agreement with those of Henkel and Joulain
concerning thick media.

\section{Summary}
Using the properties of the macroscopic field operators
appropriate for magnetodielectric dissipative systems, we have
extended our previous approach to the Casimir effect between two
dielectric multilayered stacks to magnetodielectric systems. As
expected, the resulting expression for the force between the
stacks is of the same form as for a purely dielectric system.
However, the Fresnel coefficients are modified owing to the
different magnetic properties of the system in the present case.
Using this result, we have numerically analysed the effect of the
medium dispersion on the Casimir force in a
metal-magnetodielectric system described by the Drude-Lorentz
permittivities and permeabilities. According to this analysis,
taking into account the medium dispersion does not much affect the
conclusions that one may draw from a simple nondispersive model of
Ref. \cite{Kenn02} regarding the material dependence of the force.
Our simulations also demonstrate that the force at zero
temperature in a realistic system is attractive at small distances
(covering the submicron and micron range for good metals), which
supports statements in Ref. \cite{Iann03}, and, depending on the
magnetic (dispersion) properties of the magnetodielectric, becomes
repulsive at larger distances. However, owing to contributions of
the low-frequency waves to the force, the crossover occurs at a
much smaller distance than expected from the simple $d\sim
\lambda_{Tm}$ estimate. At room temperature, the force on good
(Drude) metals such as Au, Cu, Al remains attractive over the
entire range of distances.

\section*{Acknowledgments}
The author thanks D. Iannuzzi for useful suggestions. This work
was supported by the Ministry of Science and Technology of the
Republic of Croatia under contract No. 0098001.

\end{document}